\documentclass[12pt,preprint]{aastex}

\newcommand{\etal}{{et al.~}}
\newcommand{\Mpc}{\, h^{-1} \, {\rm Mpc}} 
\newcommand{\se}{\sigma_8}
\newcommand{\vs}{v_{12}}
\newcommand{\disp}{\sigma_{12}}

\newcommand{\xbb}{\bar{\bar{\hbox{$\xi$}}}}
\newcommand{\beq}{\begin{equation}}
\newcommand{\eeq}{\end{equation}}
\newcommand{\ab}{{_{{\scriptscriptstyle A,B}}}}
\newcommand{\AB}{{_{{\scriptscriptstyle AB}}}}

\newcommand{\qab}{{q_{{\scriptscriptstyle AB}}}}
\newcommand{\qba}{{q_{{\scriptscriptstyle BA}}}}
\newcommand{\rab}{{{\vec r}\AB}}
\newcommand{\ra}{{{\vec r}_{{\scriptscriptstyle A}}}}
\newcommand{\rA}{r_{\scriptscriptstyle A}}
\newcommand{\rb}{{{\vec r}_{{\scriptscriptstyle B}}}}
\newcommand{\rh}{{\hat r\AB}}

\newcommand{\rha}{{{\hat r}_{{\scriptscriptstyle A}}}}
\newcommand{\rhb}{{{\hat r}_{{\scriptscriptstyle B}}}}
\newcommand{\sA}{{s_{{\scriptscriptstyle A}}}}
\newcommand{\sB}{{s_{{\scriptscriptstyle B}}}}
\newcommand{\va}{{{\vec v}_{{\scriptscriptstyle A}}}}

\def\degr{^\circ}

\shorttitle{$\Omega_m$ without priors}

\shortauthors{Feldman et al.}

\begin{document}

\title{An estimate of $\Omega_m$ without conventional priors}

\author{
H. Feldman$^{1,2}$, R. Juszkiewicz$^{3,4,5}$, P.
Ferreira$^{6}$, M. Davis$^{7}$, E. Gazta{\~n}aga$^{8}$,
J. Fry$^{9}$, \\
A. Jaffe$^{10}$, 
S. Chambers$^{1}$, L. da Costa$^{11}$, M. Bernardi$^{12}$,\\ 
R. Giovanelli$^{13}$, M. Haynes$^{12}$, G. Wegner$^{14}$\\ }

\bigskip
\altaffiltext1{Dept. of Physics \& Astronomy,
University of Kansas, Lawrence, KS 66045, USA}
\altaffiltext2{Racah Institute of Physics, Hebrew University, Jerusalem, Israel}
\altaffiltext3{Institute of Astronomy,
Zielona G{\'o}ra University, 
Zielona G{\'o}ra, 
Poland}
\altaffiltext4{Copernicus Astronomical Center, 
Warsaw, Poland}
\altaffiltext5{Observatoire de la C{\^o}te d'Azur, Nice, France}
\altaffiltext6{Astrophysics, University of Oxford, 
Oxford OX1 3RH, UK}
\altaffiltext7{Dept. of Astronomy, University of California,
Berkeley, CA 94720-3411, USA}
\altaffiltext8{IEEC/CSIC, 
08034 Barcelona, Spain, and INAOE, Astrofisica,
Puebla 7200, Mexico}
\altaffiltext9{Dept. of Physics, University of Florida, Gainesville
FL 32611-8440, USA}
\altaffiltext{10}{Blackett Laboratory, Imperial College, 
London SW7 2AZ, UK}
\altaffiltext{11}{European Southern Observatory, 
D-85748 Garching, Germany}
\altaffiltext{12}{Physics Dept., Carnegie Mellon University, 
Pittsburgh, PA 15213, USA}
\altaffiltext{13}{Center for Radiophysics and Space Research,
Cornell University, Ithaca, NY 14853, USA}
\altaffiltext{14}{Dept. of Physics \& Astronomy, Dartmouth College, 
Hanover, NH 03755-3528, USA}

\begin{abstract}

Using mean relative peculiar velocity measurements
for pairs of galaxies, we estimate the
cosmological density parameter $\Omega_m$ and the amplitude of density
fluctuations $\se$.  
Our results suggest that our statistic
is a robust and reproducible measure of 
the mean pairwise velocity and
thereby the $\Omega_m$ parameter.  We get $\Omega_m =
0.30^{+0.17}_{-0.07}$ and $\se = 1.13^{+0.22}_{-0.23}$.  
These estimates do not depend on prior
assumptions on the adiabaticity of the initial density fluctuations,
the ionization history, or the values of other cosmological
parameters.
\end{abstract}




\section{Introduction}

In this paper we report the culmination of a program to study cosmic
flows.  In series of recent papers we introduced a new dynamical
estimator of the $\Omega_m$ parameter, the dimensionless density of the
nonrelativistic matter in the universe.  
We use the so called {\it streaming velocity}, or the mean relative peculiar velocity for galaxy pairs, 
$\vs(r)$, where $r$ is the pair separation (Peebles 1980). It is measured directly from
peculiar velocity
surveys, without the noise-generating spatial differentiation, used
in reconstruction schemes, like POTENT (see Courteau \etal 2000 and references therein).
In the first paper of the series \citep{rj98}, we derived an equation,
relating $\vs(r)$ to $\Omega_m$ and 
the two-point correlation function of mass
density fluctuations, $\xi(r)$. 
Then, we showed that $\vs$ and $\Omega_m$
can be estimated from mock velocity surveys
\citep{pairwise1}, and finally, from real data: the
Mark III survey \citep{pairwise2}. Whenever a new statistic is introduced, 
it is of particular importance that it passes the test of 
reproducibility. Our Mark III results pass these tests: 
the $\vs(r)$ measurements are galaxy morphology- and 
distance indicator-independent.

In this {\it Letter} we extend our analysis to 
three new surveys, with the aim of testing reproducibility on a larger
sample and, in case of a positive outcome, improving on the accuracy
of our earlier measurements of $\Omega_m$ and
$\se$, the root-mean-square mass density contrast in a sphere of
radius of $ 8 \Mpc $, where $h$ is the usual Hubble parameter, $H_0\,$,
expressed in units of 100 ${\rm km \, s^{-1}Mpc^{-1}}$.  In our
notation, the symbol $\se$ always refers to matter density,
while $\se^{\rm PSCz}$ refers to the number-density of PSCz galaxies. 

Unlike our analysis, other estimators of cosmological parameters are often  
degenerate, hence $ \se$ and $ \Omega_m$ can not be extracted without making 
additional Bayesian prior assumptions, which we call {\it conventional priors}:
a particular choice of values for 
$ h$, the baryon and vacuum densities, $ \Omega_b$ and $ \Omega_{\Lambda}$, 
the character of the primeval inhomogeneities  
(adiabaticity, spectral slope, t/s ratio), the ionization history, 
etc. \citep{BLOS}. The estimates of $\Omega_m$ and $\se$
presented here {\it do not} depend on conventional priors. The only
prior assumption we make is that up to $\se$, 
the PSCz estimate of $\xi(r)$ describes the mass
correlation function. We test this assumption by comparing
the predicted $\vs(r)$ to direct observations. We also check how
robust our approach is by replacing the PSCz estimate of $\xi(r)$
with an APM estimate and two other pure power-law toy models.

\section{The pairwise motions and galaxy clustering}

\begin{figure}
\plotone{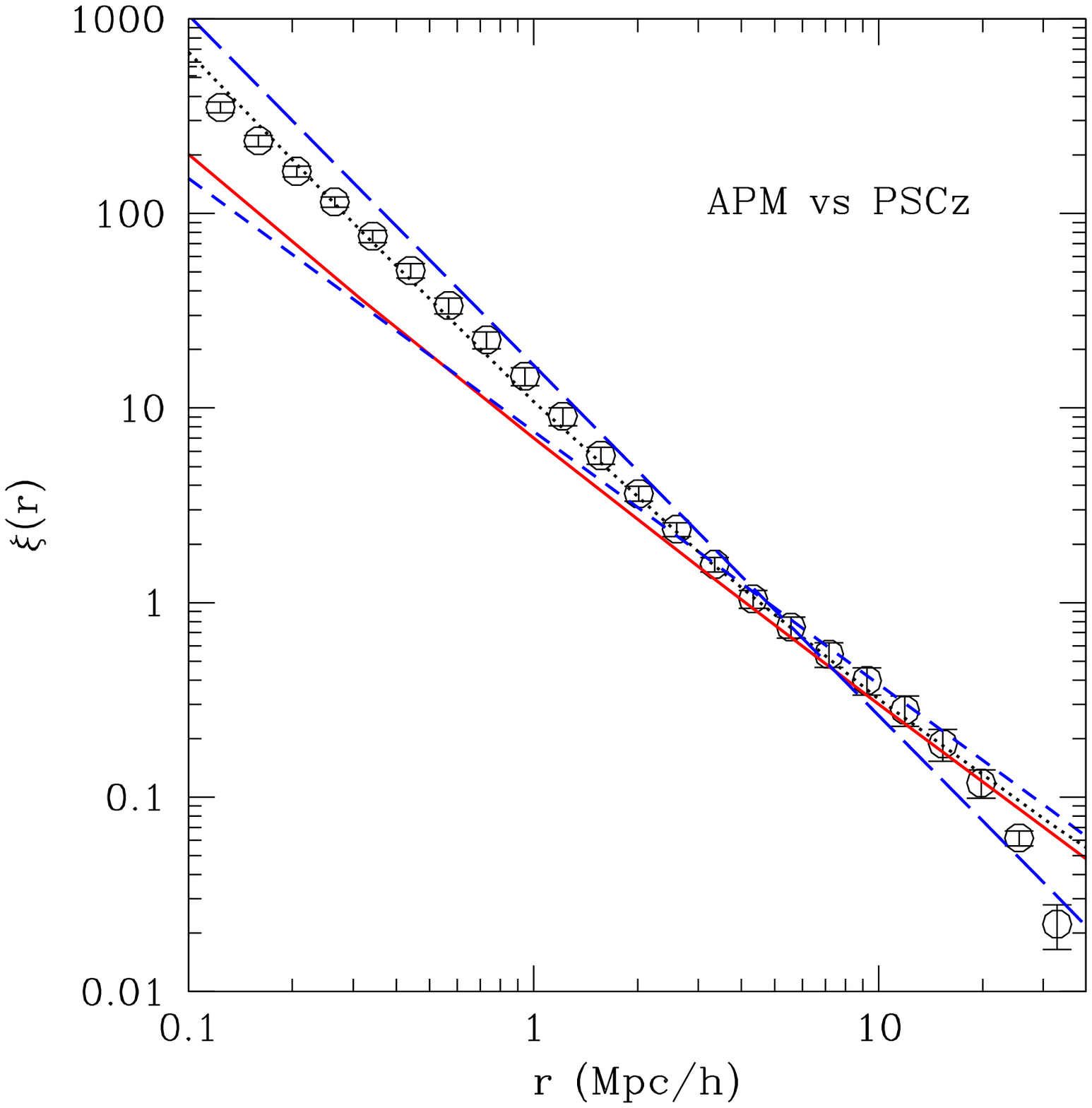}
\caption{The APM correlation function measurements (circles with
error bars) compared to four closed-form expressions for $ \xi(r)$:
two power-law toy models with slopes $ \gamma = 1.3$ (short-dashed
line) and 1.8 (long-dashed line) and two more realistic, broken
power-law empirical fits, given by eq.~[\ref{finalxi}]. The latter 
two represent the PSCz (solid line) and the APM survey (dotted line). 
All four expressions for $ \xi(r)$ assume $\se = 0.83$.
}
\label{APMfit}
\end{figure}

The approximate solution of the pair conservation equation derived by
Juszkiewicz \etal (1999) is given by
\begin{eqnarray}
\vs (r) \; &=& \; - \, {\textstyle{2\over 3}} \, H_0\,r \, \Omega_m^{0.6} 
\, \xbb (r)[1 + \alpha \; \xbb (r)] \;,
\label{21nd-order}\\
\xbb(r) \; &=& \; {3\, \int_0^r \, \xi(x) \, x^2 \, dx\over
r^3\,[\, 1 + \xi(r) \, ]} \; ,
\label{xb}
\end{eqnarray}
where $ \alpha = 1.2 - 0.65 \,\gamma $, and $ \gamma = - (d\ln
\xi/d\ln r)|_{\xi=1} $. As a model for $\xi(r)$, we
use the Fourier transform of the PSCz power spectrum (Hamilton \&
Tegmark 2002, eq.[39]), which can be expressed as
\beq
\xi(r) \; = \; (\se/0.83)^2 \; \left[ \; (r/r_1)^{-\gamma_1} 
+ (r/r_2)^{-\gamma_2}\; 
\right] \; ,
\label{finalxi}
\eeq  
where $ r_1 = 2.33 \Mpc $, $ r_2 = 3.51 \Mpc $, 
$ \gamma_1 = 1.72 $, $\gamma_2 = 1.28$,
and $\se$ is a free parameter.  If the PSCz galaxies
follow the mass distribution, then $ \se = \se^{\rm PSCz} = 0.83 $. 
The quantities $\se$ and $\xi(r)$ describe nonlinear matter
density fluctuations at redshift zero.
The PSCz fit with $\se = 0.83$ in eq.~(\ref{finalxi}) is plotted in
figure~\ref{APMfit}, together with the APM correlation function
measurements for comparison. For $\; r < 15 \Mpc$,
the APM correlation function is well approximated by
eq.~[\ref{finalxi}] with
$ r_1 = 3.0 \Mpc $, $ r_2 = 2.5 \Mpc $, $ \gamma_1 = 1.9 $ and
$ \gamma_2 = 1.1 $. 
For $ 2 \Mpc < r <15 \Mpc $, which is
the range of separations of interest here, the PSCz and
APM correlation functions in figure~\ref{APMfit}
are almost indistinguishable. This
provides an added reason to believe that choosing PSCz as a template
for $\xi(r)$ was a good idea.  To test the stability of our
conclusions with respect to uncertainties regarding the small-$r$
behavior of $\xi(r)$, we will compare predictions for $\vs(r)$
based on PSCz parameters for eq.~[\ref{finalxi}] with those
based on the APM survey.
To study the sensitivity of $\vs(r)$ and inferred cosmological
parameters to the assumed slope of
$\xi(r)$, we will also consider two simplified, pure power-law
toy models, given by
\beq
\xi(r) \; = \; (\se/0.83)^2 \; (r/r_0)^{-\gamma} \; ,
\label{pure}
\eeq
where $ \gamma = 1.3$ and 1.8, while $ r_0 = 4.76 \Mpc $ and
$ 4.6 \Mpc$, respectively.  

\section{Peculiar velocity surveys}

We will now describe our measurements.  Each redshift-distance survey
provides galaxy positions, $\ra$, and their radial peculiar
velocities, $\, \sA = \ra\cdot \va/\rA \equiv \rha\cdot \va \,$,
rather than three-dimensional velocities $\va$. We use hats to denote
unit vectors while indices $A,B = 1,2, \ldots$ count galaxies in the
catalogue. Consider a set of pairs $(A,B)$ at fixed separation $\, r =
\bigl| \rab \bigr|\,$, where $\,\rab \equiv \ra - \rb \,$. To relate
the mean radial velocity difference of a given pair to $\vs(r)$, we
have to take into account a trigonometric weighting factor,
\begin{eqnarray}
\langle \, \sA - \sB \, \rangle \; = \;
\vs(r) \; \qab \ \ \ \ \ \ \ 
\label{sAB}
\qab \; \equiv \; \rh\cdot(\rha + \rhb)/2 \; = \; - \qba \; .
\label{qAB}
\end{eqnarray}
To estimate $\vs$, we minimize the quantity
$\chi^2(\vs) \; = \; \sum_{\ab} \, \left[ (\sA - \sB) - \qab
\,\vs(r) \, \right]^2  .$
The condition $\partial \chi^2 / \,\partial\vs = 0$ implies
\begin{eqnarray}
\vs (r) \; = \; {
{\sum_{\ab} \, (\sA - \sB)\, \qab }/\; {\sum_{\ab} \qab^2}} \;\; .
\label{estimator}
\end{eqnarray}
In this study we use following independent proper distance catalogues.

1. Mark III. This survey (Willick \etal 1995, 96, 97) contains five
different types of data files: Basic Observational and Catalogue Data;
Individual Galaxy Tully-Fisher (TF) and $D_n$-$\sigma$ Distances;
Grouped Spiral Galaxy TF Distances; and Elliptical Galaxy Distances as
in the Mark II (for TF and $D_n$-$\sigma$ methods,
see Binney \& Merrifield 1998).  The subset we use here contains 2437
spiral galaxies with TF distance estimates.  The total survey depth is
over $ 120 \Mpc $, with homogeneous sky coverage up to $ 30\Mpc $.

2. SFI \citep{SFI96,SFI98,SFI99}. This is an all-sky survey,
containing 1300 late type spiral galaxies with $I$-band TF distance
estimates. The SFI catalogue, though sparser than Mark III in certain
places, covers more uniformly the volume out to $ 70 \Mpc $.

3. ENEAR \citep{ENEAR}. This sample contains 1359 early type
elliptical galaxies brighter than $m_B$ = 14.5 with $D_n$-$\sigma$
measured distances. ENEAR is a uniform, all-sky survey, probing a
volume comparable to the SFI survey.

4. RFGC \citep{RFGC}. This catalogue provides a list of radial
velocities, HI line widths, TF distances
and peculiar velocities of 1327 spiral galaxies
that was compiled from observations of flat galaxies from FGC
\citep{FGC} performed with the 305\,m telescope at Arecibo
\citep{RFGC2}. The observations are confined within the zone
$0\degr<\delta\le+38\degr$ accessible to the radio telescope.

\begin{figure}
\plotone{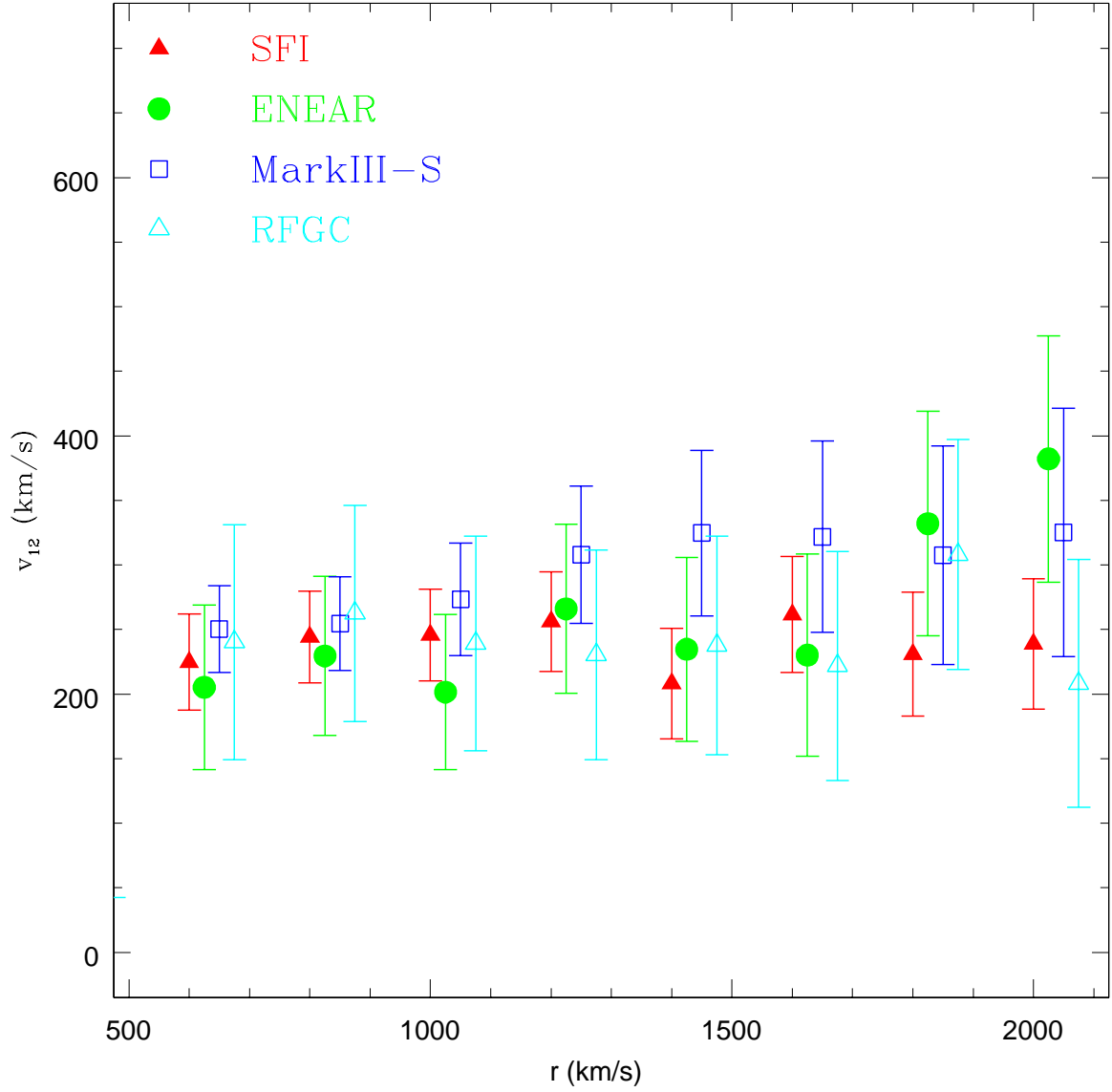}
\caption {The pairwise velocities $\vs(r)$ for the four surveys.
The Mark III-S $\vs(r)$ measurements come from our earlier work
\citep{pairwise2}.
Clearly, the results from all surveys agree well with each other.  }
\label{v12-data}
\end{figure}

Figure~\ref{v12-data} shows our estimates of $\vs(r)$.
Although the catalogues we used are independent, distinct and survey
very different galaxy and morphology types, as well as different
volumes and geometries, our results are robust and consistent with
each other. The error bars are the estimated 1-$\sigma$
uncertainties in the measurement due to lognormal distance errors
(around $15\%$), sparse sampling (shot noise), and finite volume 
of the sample (cosmic variance). For more details
on error estimates used here, see \citet{landy92}, Haynes \etal
(1999a,b), and \citet{pairwise1}.  

The agreement among the $\vs(r)$ estimates from different surveys, 
plotted in figure~\ref{v12-data} becomes 
even more impressive when compared to discrepancies between  
different estimates of a close cousin of our
statistic, the pairwise velocity dispersion, $\disp(r)$.
The velocity dispersion appears to be less
sensitive to the value of $\Omega_m$ than to the presence of rare,
rich clusters in the catalogue and to galaxy morphology, with
estimates of $\disp$ at separations from one to a few Mpc varying from
300 to $ 800 \, \rm km \, s^{-1} $ from one survey to another
\citep{DP1983,ZQSW,marzke,ZJB}.
The lack of systematic differences 
between $\vs(r)$ estimates in figure~\ref{v12-data} is incompatible
with the linear
biasing theory unless the relative elliptical-to-spiral bias,
$b_{\rm E}/b_{\rm S}$, is close to unity at separations $r > 5 \Mpc$,
in agreement with our earlier studies \citep{pairwise2}; for the same
reason our results strongly disagree with recent
semi-analytic simulations \citep{sheth, jing}. 

\begin{figure}
\plotone{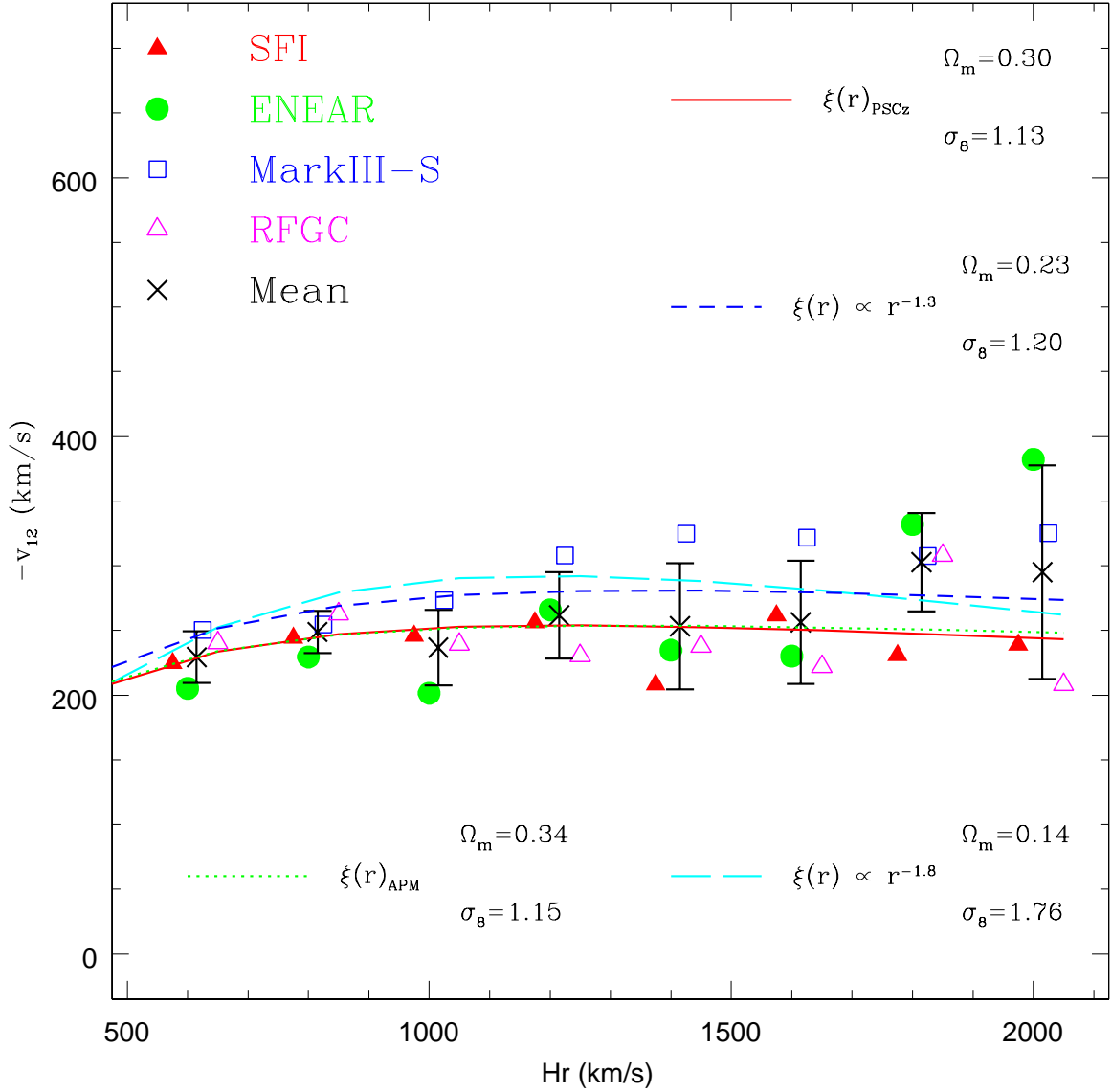}
\caption {The crosses and the associated error bars show 
the weighted mean pairwise velocity, obtained by averaging over four
surveys. Individual survey data points are also shown; we have
suppressed their error bars for clarity. These direct measurements
of $ \vs$ are compared to four $ \vs(r)$ curves, derived by assuming 
four different models of  $ \xi(r)$, plotted in figure 1. The labels 
identify best fit $ \Omega_m$ and $ \se$ parameters.}
\label{v12-all}
\end{figure}

In figure~\ref{v12-all} we show the results for each of the catalogues
we investigated, as in figure~\ref{v12-data},
but now we overlay the weighted mean of the individual
catalogues. Since the results are robust, combining the catalogues
reduces the errors and gives us a strong prediction for the parameter
values. Figure~\ref{v12-all} shows the results of our theoretical best fits: 
the solid (dotted) line follows
the double power law correlation function using the PSCz (APM)
correlation function (eq.~[\ref{finalxi}]). 
Clearly, the slope
differences in $\xi(r)$ at small separations do not affect $\vs(r)$
in the range of separations we consider. Moreover, given the error
bars on $\vs$, the $\gamma = 1.3$ power-law toy model prediction
for $\vs(r)$,
as well as the resulting best fit values of $ \se $ and $ \Omega_m $  
are similar to those based on the APM and PSCz correlation functions.
For $ \se \approx 1$ and $ \xi(r) \propto r^{-\gamma}$ at 
$ r > 10 \Mpc$, linear theory applies and
$ \vs \propto r^{1 - \gamma}$. Therefore all
three of the models considered above give $\vs \propto r^{-0.3}$, in
good agreement with the observed nearly flat $\vs(r)$ curve. 
All of the above does not apply to our $\gamma = 1.8$ toy model,
which is significantly steeper than the APM and PSCz $ \xi(r)$ at 
large $r$, and for $ \se \approx 1$,
the $\vs(r)$ is expected to drop almost by half between 10 and
$ 20 \Mpc$. It is possible to flatten the $ \vs(r)$ curve
only by increasing $ \se$ and extending the nonlinear regime 
to larger separations. The example considered here gives 
$ \se = 1.76$, in conflict with all other
estimates of this parameter (see the discussion below). 
Correlation functions, steeper than APM or PSCz often appear
in semi-analytic simulations and this example shows how 
$\vs(r)$ measurements can be used to constrain those models.
 
\begin{figure}
\plotone{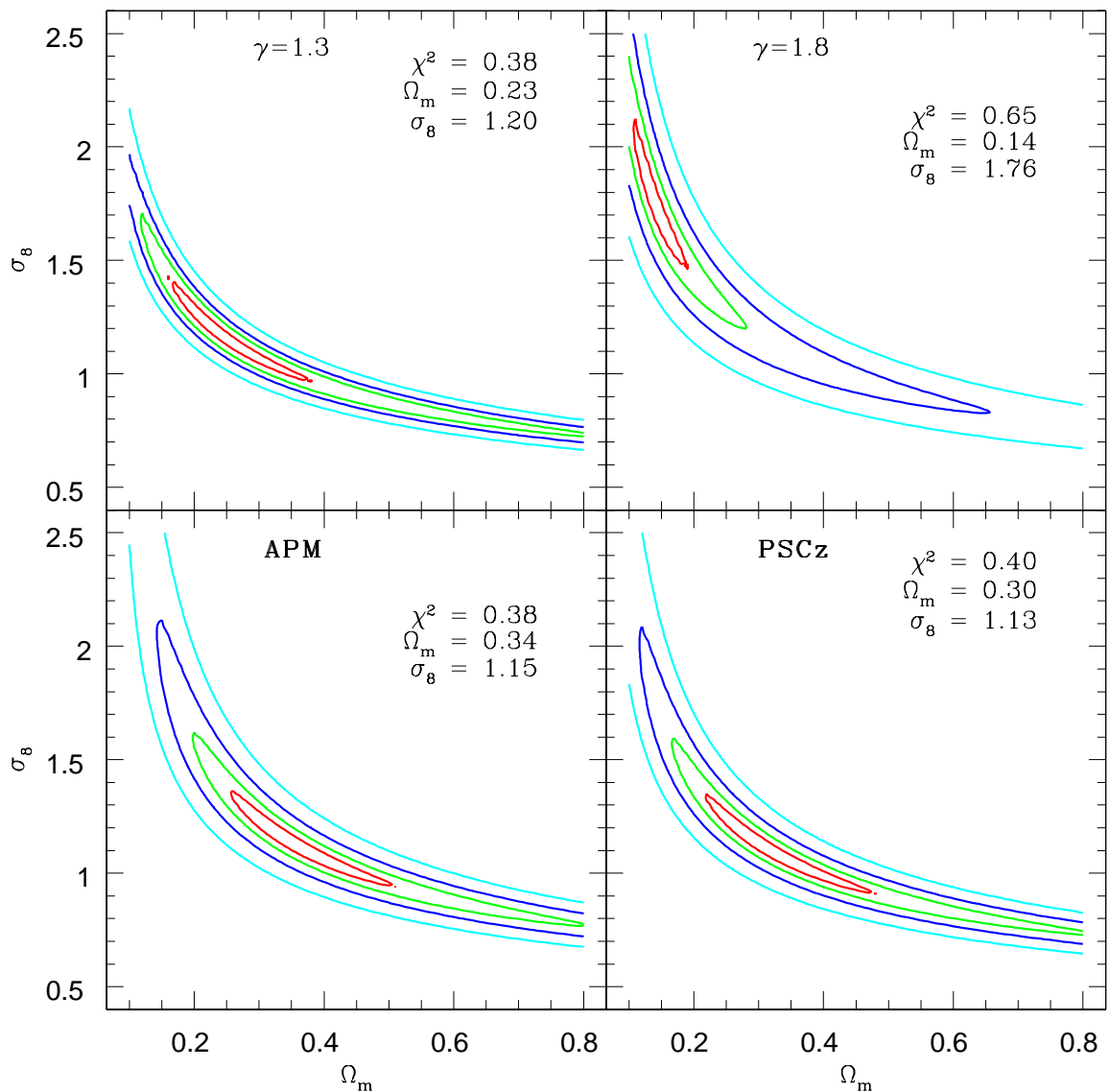}
\caption {The results of the maximum likelihood analysis. The 
upper panels show results for power-law toy models, while the
bottom panels are based on realistic representations of 
observations: the APM and PSCz data, respectively. 
Likelihood peak coordinates and the values of $ \chi^2$ for 
each model are also indicated. The innermost contours define 
the 68\%, or 1-$\sigma$ areas around the peaks.  The remaining 
nested contours show the 2, 3 and 4-$\sigma$ boundaries.}
\label{contplot}
\end{figure}

In figure~\ref{contplot} we plot the
resulting 1,2,3 and $4\sigma$ likelihood contours in the $(\Omega_m,
\se)$ plane.  
The quoted errors define the 1$\sigma$, or 68\% statistical
significance ranges in each of the two parameters and correspond to
the innermost contour in figure~\ref{contplot}. 
The low $\chi^2$ per degree of freedom is indicative of the
correlations between $\vs(r)$ measurements at different separations
$r$. One of the sources of correlations is the finite depth of our 
surveys. Note also that since we are dealing with pairs of galaxies,
the same galaxy can in principle influence all separation bins. 
The contours derived using the PSCz correlation function
(eq.~[\ref{finalxi}]), are shown in the bottom right panel. 
The best fit values are
\beq
\Omega_m \, = \, 0.30^{+0.17}_{-0.07} \;\;\; {\rm and} \;\;\;
\se \, = \, 1.13^{+0.22}_{-0.23} \; .
\label{best}
\eeq
The likelihood
contours based on the APM correlation function (with best
fit values $\, \Omega_m = 
0.34^{+0.16}_{-0.14}\,$ and $\, \se = 1.15^{+0.15}_{-0.20}$) and
the $ \gamma = 1.3$ power-law model ($\, \Omega_m = 
0.23^{+0.15}_{-0.06}\,$ and $\, \se = 1.20^{+0.20}_{-0.25}$)
are similar. 
Our estimate of $\se$ agrees with the results of studies of clustering
of galaxy triplets in real and Fourier space in three different
surveys: the APM \citep{eg95,fg99}, the PSCz \citep{FFFS01} and the
2dF \citep{verde01}.  A similar value of $\se$ was recently inferred
from the observed position of the inflection point in the APM $\xi(r)$
\citep{gj01}. All of the above measurements are consistent with a
$\se$ within 20\% of unity.  A $\se$ close to unity follows from
maximum likelihood analysis of weak gravitational lensing
\citep{Wetal02} after assuming $ \Omega_{\Lambda} = 0.7$,
$ \Omega_m = 0.3$ and $h= 0.7$.  Measurements of the abundance of clusters
\citep{Betal02} tend to give $\se$ closer to the lower end of our 68\%
interval if $\Omega_m = 0.3$. The good agreement between these
results, obtained with different methods, riddled with systematic
errors of different nature, suggests that our estimates of statistical
errors are reasonable and that the systematic errors are subdominant
(unless there is an evil cosmic conspiracy of errors).
The parameters in eq.~[\ref{best}]
also agree with those inferred from the power spectrum of
the anisotropy of the cosmic microwave background (CMB) temperature
distribution on the sky: $\se = 0.9 \pm 0.1$, and $\Omega_m = 0.29 \pm
0.07$ (see Table 2 in Spergel \etal 2003).  It is important to bear in
mind, however, that unlike the CMB results, our 
estimates were obtained from the
velocity and PSCz data alone, without the conventional
priors. Therefore the
$\vs(r)$ measurements combined with the CMB or the supernova data can
be used to break the cosmological parameter degeneracy.
Choosing $ \gamma = 1.8$, which is significantly
steeper than the observed $ \xi(r)$, gives 
$ \Omega_m = 0.14^{+0.06}_{-0.04}$ and $ \se = 1.76^{+0.34}_{-0.26}$
(figure 4, upper right),
in conflict with all of the independent estimates of $\se$,
discussed above. This suggests that the observed slope of the
APM and PSCz correlation functions is close to the slope of
the dark matter correlation function.

\null

\noindent{\bf Acknowledgments:} 
HAF wishes to acknowledge support from the NSF
under grant number AST--0070702, the University of Kansas General
Research Fund, the National Center for Supercomputing Applications, 
the Lady Davis Foundation and the Schonbrunn Fund at the Hebrew
University, Jerusalem and by the Institute of Theoretical Physics at
the Technion, Haifa, Israel. RJ wishes to thank
Uriel Frisch for his hospitality at The Observatoire de la
C{\^o}te d'Azur and also acknowledge support by a KBN 
grant 2P03D01719 (Poland),  
the Tomalla Foundation (Switzerland) and the Rose Morgan
Visiting Professorship at the University of Kansas. PGF thanks the
Royal Society. EG acknowledges support from INAOE, the
Spanish MCyT, project AYA2002-00850 and EC-FEDER funding. 
This work began at the 1997 Summer Workshop at the
Aspen Center for Physics, and we thank the Organizers.



\end{document}